\documentclass{aa}

\usepackage{graphicx}
\usepackage{hhline}
\usepackage{array}
\usepackage{float}
\usepackage{amsmath}
\usepackage{rotating}

\begin{document}

\title{The $N$-point correlation functions of the \emph{COBE}-DMR maps
revisited} 

\author{Hans K. Eriksen\inst{1} \and A. J. Banday\inst{2} \and
Krzysztof M. G\'orski\inst{3,4} }

\institute{
Institute of Theoretical Astrophysics, University of Oslo, P.O. Box
1029 Blindern, N-0315, Norway 
\and 
Max-Planck-Institut f\"ur Astrophysik, Garching bei M\"unchen, Germany 
\and 
European Southern Observatory, Garching bei M\"unchen, Germany
\and
Warsaw University Observatory, Aleje Ujazdowskie 4, 00-478 Warszawa, Poland
}

\date{Recieved 19 June 2002 / Accepted 28 August 2002}

\abstract{We calculate the two-, three- and (for the first time) 
four-point correlation functions of the \emph{COBE}-DMR 4-year sky
maps, and search for evidence of non-Gaussianity by comparing the 
data to Monte Carlo-simulations of the functions. The analysis is performed for 
the 53 and 90 GHz channels, and five linear combinations thereof. 
For each map, we simulate an ensemble of 10\,000 Gaussian realizations 
based on an \emph{a priori} best-fit scale-invariant cosmological power spectrum, 
the DMR beam pattern and instrument-specific noise properties.
Each observed \emph{COBE}-DMR map is compared to the ensemble using a 
simple $\chi^2$ statistic, itself calibrated by simulations.
In addition, under the assumption of Gaussian fluctuations, we find explicit
expressions for the expected values of the four-point functions in
terms of combinations of products of the two-point functions,
then compare the observed four-point statistics to those predicted by 
the observed two-point function, using a redefined $\chi^2$ statistic.
Both tests accept the hypothesis that the DMR maps are consistent
with Gaussian initial perturbations. 
}

\maketitle

\section{Introduction}

The study of CMB temperature anisotropies and their statistical
properties has become an important theme in modern cosmology. In its
most conventional interpretation, the distribution of anisotropies
reflects the properties of the universe 
approximately 300\,000 years after the Big Bang, at the surface of last
scattering. Thus, by measuring statistical quantities such as the angular
power spectrum or the angular two-point correlation function, we can
infer the values of many interesting cosmological parameters. 

For both theoretical and practical purposes, it is convenient to
expand the temperature anisotropy field into a sum of (complex) 
spherical harmonics: 
\begin{equation}
\Delta T(\theta, \phi) = \sum_{lm} a_{lm} \: Y_{lm}(\theta, \phi)
\end{equation}
The temperature perturbation field is said to be Gaussian distributed
if each $a_{lm}$ follows an independent Gaussian probability
distribution. The question of whether the observed temperature field
is Gaussian or otherwise is of crucial importance for modern
cosmology. From a scientific standpoint, most conventional
inflationary models of structure formation predict a Gaussian temperature
field, whereas scenarios which invoke topological defects to
seed the large-scale structure predict a non-Gaussian distribution. 
Thus the statistical properties of the $a_{lm}$'s can be used to distinguish
such models. Secondly, from an analysis point-of-view, most parameter 
estimation techniques -- usually based on the observed angular power
spectrum -- assume Gaussianity, and may therefore be biased if the observed 
field is indeed non-Gaussian. 

However, testing for non-Gaussianity is anything but trivial, and several
qualitatively different tests are required in order to perform a
complete analysis. At present the arsenal of available tests which
have been applied to the \emph{COBE}-DMR data consists of
at least the following: bi- and trispectrum based analysis 
(Ferreira et al. \cite{ferreira}; Magueijo \cite{magueijo};
Sandvik \& Magueijo \cite{sandvik}; Komatsu et al. \cite{komatsu};
Kunz et al. \cite{kunz}), 3-point correlation function based tests
(Kogut et al. \cite{kogut}), methods utilizing wavelets (Cay\'on et
al. \cite{cayon}; Barreiro et al. \cite{barreiro})  
and Minkowski functionals (Schmalzing \& G\'orski \cite{schmalzing98};
Novikov et al. \cite{novikov}).
Indeed, there has been a small resurgence in interest in the
possibility of non-Gaussian signals in the \emph{COBE}-DMR
maps as a consequence of the bispectrum work of Ferreira 
et al. (\cite{ferreira}) and Magueijo (\cite{magueijo}). 
These papers find non-Gaussian contributions using
harmonic analyses at the 98\% confidence limit, and 
although Banday et al. (\cite{bandaya}) explain these tentative
detections by appealing to the presence of a specific residual
systematic artifact in the data, additional investigation is
warranted.

In this paper we adopt $N$-point correlation functions as 
probes for non-Gaussianity. For a Gaussian field all odd $N$-point
functions (such as the three-point function) have vanishing
expectation values, while all even $N$-point functions can be reduced
to expressions involving the two-point function. 
Thus, if the observed three-point function
is significantly non-zero when compared to a Gaussian ensemble, its native
distribution is probably non-Gaussian. Further, if the four-point
function does not reduce into two-point functions, the same conclusion
can be made.

The first part of this paper builds on ideas demonstrated in 
Kogut et al. (\cite{kogut}) and Hinshaw et al. (\cite{hinshaw}). 
We study the 4-year \emph{COBE}-DMR sky maps, computing the two- and
three-point functions, as has been performed previously, then
proceeding to extend the analysis for the first time to the 
determination of several four-point functions. 
The definitions of these new functions are given in
Sect.\ \ref{sec:definitions}. 

In Sect.\ \ref{sec:measurements} we compute the  
various correlation functions for the four DMR channels
and five linear combinations thereof.
Next, we compute the same functions for 10\,000 Monte Carlo simulated
Gaussian maps, which are used as the basis of the various statistical
tests of non-Gaussianity. Initially, we apply the $\chi^2$ test as
defined by Kogut et al. \cite{kogut}, by comparing the observed data
value to the distribution generated from the application
of the $\chi^2$ statistic for each map in the simulated ensemble.

Subsequently, in Sect.\ \ref{sec:reduction}, we provide
expressions for the expected value of several four-point functions
in terms of the two-point function, then explicitly 
compare the observed four-point functions to those predicted by the 
observed two-point function. We define a suitable $\chi^2$ statistic 
in order to quantitatively measure the degree of deviation, once again
calibrated by Monte Carlo-simulations.

\section{Definitions of the correlation functions}
\label{sec:definitions}


An $N$-point correlation function is defined as the average product of
$N$ temperatures with \emph{a fixed relative orientation} on the sky:
\begin{equation}
C_{N}(\theta_{1}, \ldots, \theta_{2N-3}) = \biggl<T(\hat{n}_{1}) \,
T(\hat{n}_{2}) \cdots T(\hat{n}_{N}) \biggr>
\end{equation}
where $\hat{n}_i$ is the direction vector of the $i$'th pixel in the
configuration, and 
$\cos\theta_j = \hat{n}_k\cdot\hat{n}_l$ for $2N-3$
arbitrary, but different, angular pixel distances on the sky.  

Although the $N$-point functions are easily defined and relatively
simple to implement computationally, their evaluation is generally
CPU-intensive, which is especially problematic since detailed
assessment of results requires large Monte Carlo simulation data sets.
The full computation of an $N$-point correlation function scales as
$\mathcal{O}(N_{\mathrm{pix}}^N)$, and is therefore virtually impossible to
compute for high-resolution maps for any order $N$ greater than
two. For this reason we choose to compute only a 
subset of the possibilities in the $N$-dimensional configuration
space, designed to reduce the complexity of the problem. As an example 
consider the pseudo-collapsed three-point function for which we
require two points to coincide, effectively reducing the geometry to 
that of the two-point function. Such subsets typically scale somewhere
between $\mathcal{O}(N_{\mathrm{pix}}^2)$ and $\mathcal{O}(N_{\mathrm{pix}}^3)$. Thus,
with some effort put into the implementation these functions can be
computed even for rather high-resolution maps. 

Previous work has considered two special three-point functions, namely
the collapsed and the equilateral functions (Kogut et al. \cite{kogut};
Hinshaw et al. \cite{hinshaw}). As mentioned above, the collapsed
function is defined by requiring two of the three point to coincide,
while the equilateral function requires the three points to span an
equilateral triangle on the sphere. 

In this paper, we shall also consider several simple four-point
configurations. These functions are, in order of complexity: 
\begin{enumerate}
\item the collapsed 1+3 point function 
\item the collapsed 2+2 point function 
\item the collapsed equilateral four-point function 
\item the rhombic four-point function 
\end{enumerate}
The names should be self-explanatory. For the 1+3 point
function, three points coincide in a manner similar to the collapsed 
three-point function. The 2+2 point function is defined by allowing two
pixel pairs to coincide. The collapsed equilateral function is the
equilateral three-point configuration where one pixel is multiplied
twice. The last case, the rhombic four-point function, consists of two
equilateral triangles ``glued'' together along one side, effectively
spanning a rhombus on the sphere. Note that these functions are chosen
because of ease of implementation, not because they are
better suited for the testing of Gaussianity than other configurations. 

Several of the functions defined above are so-called collapsed
functions, i.e. one pixel is multiplied one or more times with itself.
Unfortunately, for noisy maps this renders the function completely 
noise dominated. 
To remedy this problem we substitute the collapsed 
functions by so-called pseudo-collapsed versions, as introduced by Hinshaw et
al. (\cite{hinshaw}).  
For the \emph{COBE}-DMR experiment the beam size is approximately
$7\degr$, while the pixel size is -- necessarily for adequate
sampling -- $\sim 1.8\degr$ (for the HEALPix
$N_{\mathrm{side}} = 32$
pixelization used here). Therefore the CMB signal component 
between two neighboring pixels is highly coherent, whereas 
the noise contributions are independent. Thus, instead of multiplying
a given pixel by itself several times, we multiply the pixel by one or more
of its immediate neighbors, then sum over all such possible products,
effectively multiplying by an average over the nearest
neighbors. 
Hence, we more generally define a pseudo-collapsed function as an average
product of pixels where at least one pixel is multiplied in the
pseudo-collapsed sense, ie.\ by an average over its neighbors. 
The golden rule for our analysis is that \emph{no pixel is 
ever multiplied with itself}. This definition is then not completely
equivalent to that introduced by Hinshaw et al. (\cite{hinshaw}) They 
defined the pseudo-collapsed function as the average product 
of 1) a center pixel, 2) one of its neighbors and 3) a far point,
where the far point was not allowed to be the center pixel. However,
it was allowed to be the neighboring pixel. Although not a major
problem for the three-point function, we have determined that 
the inclusion of such a product renders the first bin of the
four-point functions completely noise dominated.

\begin{figure*}
\resizebox{\hsize}{!}{\includegraphics{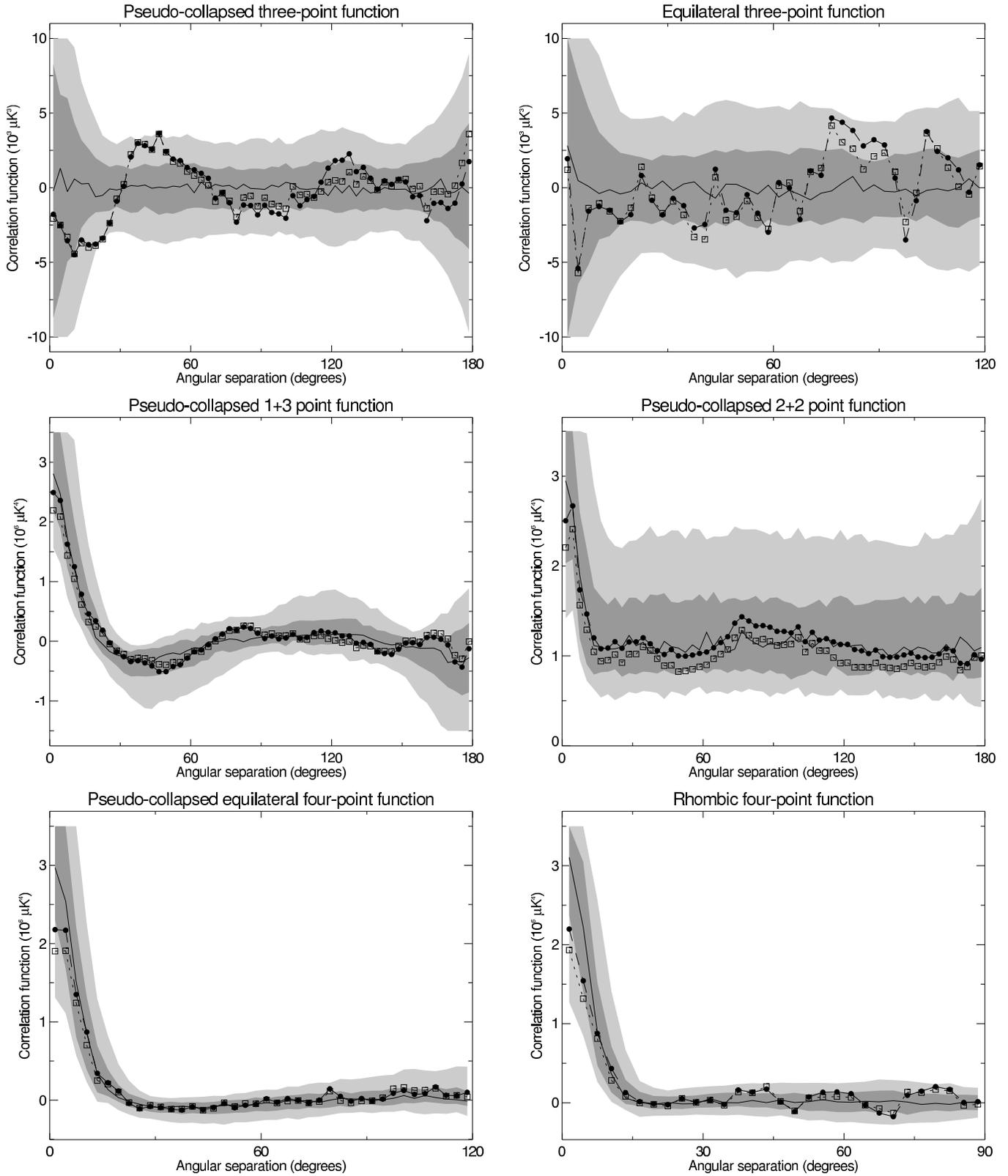}}
\caption{Three- and four-point correlation functions of the co-added
4-year DMR map. Solid line shows the most likely value for each
bin, dark shading shows the 68\% confidence region and light shading
the 95\% confidence region, as computed by Monte
Carlo-simulations. Dots represent functions for the uncorrected
co-added map, and boxes shows the functions for the map for which
high-latitude Galactic emission has been removed. Note the different
angular units on the horizontal axis, reflecting the fact that the
various functions are defined on different angular intervals.}
\label{fig:corrfuncs}
\end{figure*}

We also introduce one further small change compared to Hinshaw et
al. (\cite{hinshaw}) in that we exclude the zeroth angular bin (for
which all $N$ pixels coincide) as any cosmological information here is
heavily suppressed due to the low signal-to-noise ratio. Indeed, the
inclusion of the zeroth bin only acts to increase the
variance of the $\chi^2$ statistic, and is therefore better omitted.

\section{Measurements of the \emph{COBE}-DMR correlation functions}
\label{sec:measurements}

The \emph{COBE}-DMR experiment resulted in six independent maps, two for each
of the three frequencies at 31.5, 53 and 90 GHz. In this work
we only include the maps from the 53 and 90 GHz channels, as they
are superior in terms of the signal-to-noise ratio. 
The maps are analyzed in the 
HEALPix\footnote{\emph{http://www.eso.org/science/healpix/}} 
pixelization 
scheme, with a resolution parameter of $N_{\mathrm{side}} = 32$,
corresponding to 12\,288 pixels on the sky.
At each frequency we compute the \lq sum' $(A+B)/2$ and
\lq difference' $(A-B)/2$ combinations, which yield, respectively,
maps with enhanced signal-to-noise or noise content alone.
In addition, we also generate a co-added map from the four basic 
channels using weights to achieve the optimal signal-to-noise ratio.

The two-, three- and four-point correlation functions for these nine
map combinations are then computed. All pixels corresponding to the extended 
Galactic cut (Banday et al. \cite{banday} but recomputed explicitly for the
HEALPix scheme) are rejected from the analysis, leaving a total of 7880 accepted
pixels. The best-fitting monopole, dipole and quadrupole are
subtracted from each map before the $N$-point functions are evaluated.
The observed correlation functions are also computed
after correction for the diffuse foreground emission at high Galactic
latitude, using information from the (appropriately scaled) 
DIRBE $140\ \mu \mathrm{m}$ map (G\'orski et al. \cite{gorski}). 

For our Monte Carlo ensemble, we simulate 10\,000 individual
realizations of the CMB sky, based on an \emph{a priori} best-fit 
cosmological power spectrum. In particular, we consider
scale-invariant Gaussian temperature fluctuations ($P(k) \propto Q_{\mathrm{rms-PS}}^2
k^n$ with $n=1$) with $Q_{\mathrm{rms-PS}} = 18\ \mu \mathrm{K}$,
(G\'orski et al. \cite{gorski}). The power-spectrum is filtered through the DMR
beam and pixel window functions.  
To each simulated CMB sky, we add four noise realizations
based on the rms noise levels and observation patterns of the
observed 53 and 90 GHz sky maps. These are then combined to generate
the corresponding sum, difference and co-added sky maps.
These are then processed in an identical fashion to the DMR data.

We note that we have also assumed that there are no significant 
pixel-pixel noise correlations, although
some will indeed be present as a consequence of the differential nature
of the radiometers which couple observations separated by $\sim
60\degr$ on the sky. Lineweaver et al. (\cite{lineweaver94})
have investigated this effect in detail, and find a small
excess signal is present in the 2-point correlation function at 
$60\degr$ for maps containing noise signal alone. However, we do not
expect our results to be compromised by this assumption.

Fig.\ \ref{fig:corrfuncs} shows the results from these calculations
for the co-added maps.  The observed functions
lie comfortably within the confidence region defined by the Monte
Carlo-simulations and there are no striking deviations
visible by simple inspection. 

\section{The $\chi^2$ statistic}
\label{sec:chisquare}

In order to quantitatively measure the agreement between the DMR maps
and the simulated ensemble, we utilize the same $\chi^2$ methodology 
described by Kogut et al. (\cite{kogut}):
\begin{equation}
\chi^2 = \sum_{\alpha \beta} (D_{\alpha} - \bigl<S_{\alpha}\bigr>)
(\mathbf{M}^{-1})_{\alpha \beta} (D_{\beta} - \bigl<S_{\beta}\bigr>)
\label{eq:chisq}
\end{equation}
Here $\alpha$ and $\beta$ denote angular bins, $D$ the DMR
correlation function, and $\bigl<S_{\alpha}\bigr>$ the mean function computed
from simulations. $\mathbf{M}$ is the binned covariance matrix:
\begin{equation}
\mathbf{M}_{\alpha \beta} = \frac{1}{N} \sum_{i} (S^{i}_{\alpha}
- \bigr<S_{\alpha}\bigl>) (S^{i}_{\beta} - \bigl<S_{\beta}\bigr>)
\label{eq:m_mat}
\end{equation}

The probability density function of this $\chi^2$ statistic is
established by computing the same statistic for all maps in the ensemble,
(ie.\ by substituting $D$ with each of the simulated maps, $S^{i}$). 
The resulting histogram represents the probability distribution function against
which we compare the values from the DMR maps. 
Table \ref{tab:chisq} records 
the fraction of simulated maps with \emph{higher} $\chi^2$ values than the
observed DMR map. Thus, values of order 0.01 or 0.99 can be considered 
suspicious, while anything from 0.05 to 0.95 is acceptable. 

\begin{table*}[!t]
\centering
\begin{tabular}{|c|l|cccc|cccc|c|}\hhline{*{11}{-}}  
\multicolumn{2}{|c|}{}&\multicolumn{4}{c|}{53 GHz maps}& \multicolumn{4}{c|}{90 GHz maps} &
\\ \hhline{~~*{9}{-}}
\multicolumn{2}{|c|}{}&$A$& $B$ & $\frac{\mathrm{A+B}}{2}$ & $\frac{\mathrm{A-B}}{2}$&$A$& $B$ &
$\frac{\mathrm{A+B}}{2}$ & $\frac{\mathrm{A-B}}{2}$ & Co-added \\ \hline
&Two-point 				& 0.48 & 0.22 & 0.37 & 0.32 &
0.05 & 0.48 & 0.67 & 0.02 & 0.63 \\ 
&Pseudo-collapsed three-point		& 0.13 & 0.33 & 0.65 & 0.42 &
0.51 & 0.05 & 0.43 & 0.27 & 0.59 \\ 
&Equilateral three-point		& 0.20 & 0.84 & 0.29 & 0.65 &
0.76 & 0.95 & 0.92 & 0.49 & 0.67 \\ 
&Pseudo-collapsed 1+3 point		& 0.64 & 0.59 & 0.62 & 0.17 &
0.78 & 0.83 & 0.85 & 0.53 & 0.81 \\ 
&Pseudo-collapsed 2+2 point  		& 0.47 & 0.62 & 0.61 & 0.96 &
0.29 & 0.34 & 0.94 & 0.16 & 0.79 \\ 
\hspace*{1mm}
\begin{rotate}{90}
Uncorrected
\end{rotate}
&Pseudo-collapsed equilateral four-point & 0.36 & 0.71 & 0.70 & 0.79 &
0.95 & 0.61 & 0.57 & 0.33 & 0.92 \\ 
&Rhombic four-point			& 0.58 & 0.99 & 0.50 & 0.17 &
0.49 & 0.96 & 0.76 & 0.10 & 0.80 \\ 
&All functions combined  		& 0.34 & 0.56 & 0.43 & 0.45 &
0.45 & 0.28 & 0.79 & 0.05 & 0.75 \\ \hline

&Two-point 				& 0.52 & 0.25 & 0.36 & 0.32 &
0.05 & 0.49 & 0.63 & 0.02 & 0.60 \\ 
&Pseudo-collapsed three-point		& 0.19 & 0.41 & 0.83 & 0.42 &
0.64 & 0.09 & 0.53 & 0.27 & 0.74 \\ 
&Equilateral three-point		& 0.34 & 0.80 & 0.43 & 0.65 &
0.82 & 0.97 & 0.93 & 0.49 & 0.84 \\ 
&Pseudo-collapsed 1+3 point		& 0.67 & 0.63 & 0.65 & 0.17 &
0.81 & 0.88 & 0.65 & 0.53 & 0.82 \\ 
&Pseudo-collapsed 2+2 point  		& 0.55 & 0.50 & 0.62 & 0.96 &
0.41 & 0.40 & 0.95 & 0.16 & 0.79 \\ 
\hspace*{1mm}
\begin{rotate}{90}
Corrected
\end{rotate}
&Pseudo-collapsed equilateral four-point & 0.49 & 0.77 & 0.83 & 0.79 &
0.97 & 0.78 & 0.74 & 0.33 & 0.95 \\ 
&Rhombic four-point			& 0.53 & 0.99 & 0.52 & 0.17 &
0.58 & 0.96 & 0.78 & 0.10 & 0.88 \\ 
&All functions combined  		& 0.45 & 0.59 & 0.54 & 0.45 &
0.58 & 0.39 & 0.80 & 0.05 & 0.70 \\ \hline
\end{tabular}
\caption{Results from $\chi^2$ tests. The numbers indicate the fraction of
simulated realizations with $\chi^2$ value \emph{higher} than for the
respective \emph{COBE} map. The lower half shows the results for the 
DMR maps after correction for high latitude Galactic emission.
The upper half shows the results for the uncorrected maps.
The effect of Galactic emission appears to be minimal: this is not
unexpected since a best-fit quadrupole has been removed from the data
before analysis, and the Galactic emission is dominated by such
large-scale structure.} 
\label{tab:chisq}
\end{table*}

In Kogut et al. (\cite{kogut}) the results for the pseudo-collapsed and the equilateral
three-point functions are given for the 53 GHz $(A+B)/2$ map; they
find the fractions to be respectively 0.66 and 0.31, while we find
0.65 and 0.29. Considering the minor changes in the definitions of the
correlation functions and the different pixelizations used, the
agreement is most satisfactory.

Overall, the numbers indicate that the \emph{COBE}-DMR 
maps agree very well with the simulations. 
The optimal co-added map, for which the signal-to-noise ratio is the
highest, returns results comfortably in the accepted range,
as does the combined analysis of all $N$-point functions.
We conclude that the DMR maps are compatible with the Gaussian hypothesis
as measured by this test.  

\section{Reducing four-point functions into two-point functions}
\label{sec:reduction}

\begin{figure*}
\resizebox{\hsize}{!}{\includegraphics{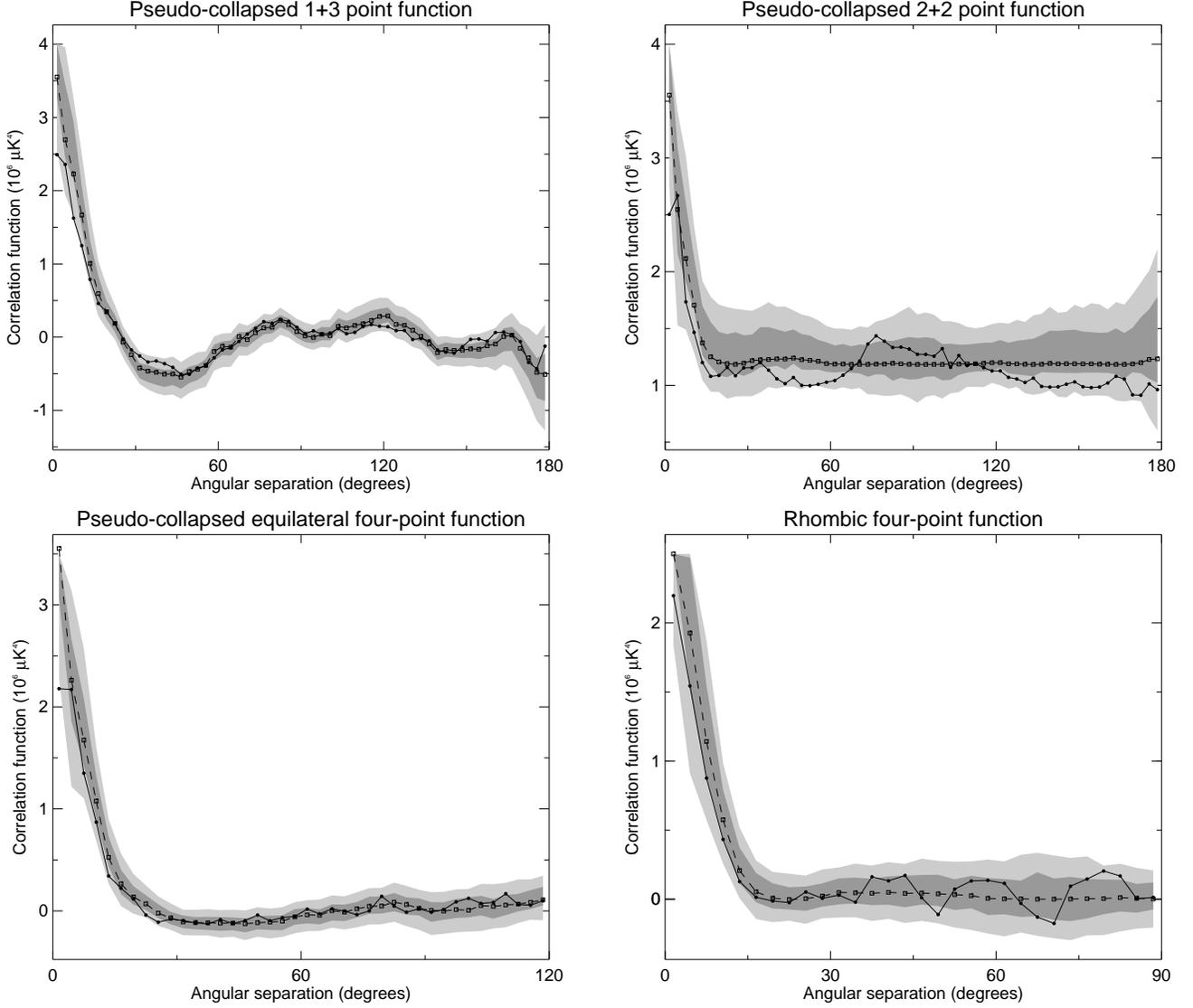}}
\caption{Comparison of the observed and the ``reduced'' four-point
functions for the co-added DMR sky map. Boxes indicate the function
predicted by the two-point functions, while shaded areas represent
the 68\% and 95\% confidence intervals computed from Monte
Carlo-simulations. The observed four-point functions are shown with a
solid line.}
\label{fig:reduced}
\end{figure*}
Since it may be noted that all even-ordered $N$-point functions have
non-vanishing expectation values determined by the two-point function, 
we can define an additional test of Gaussianity for the four-point functions. 
Explicitly, we take advantage of the following property
(as, for example,  described in Adler \cite{adler}): 
If $Y_{1}, Y_{2}, \ldots, Y_{n}$ is a
set of real-valued random variables having a joint Gaussian
distribution and zero means, then for any integer $m$: 
\begin{gather}
\bigl<Y_{1}Y_{2} \cdots Y_{2m+1}\bigr> = 0 \\
\bigl<Y_{1} Y_{2} \cdots Y_{2m}\bigr> = \sum \bigl<Y_{i_{1}} Y_{i_{2}}\bigr> \cdots
\bigl<Y_{i_{2m-1}} Y_{i_{2m}}\bigr>
\end{gather}
Here the sum goes over all $(2m-1)!!$ different ways of grouping
$Y_{1}, Y_{2}, \ldots, Y_{2m}$ into $m$ pairs. 

Thus, if the CMB temperature anisotropy field is in fact Gaussian distributed
with zero mean, then all even $N$-point functions can be reduced to combinations
of products of the two-point function. In particular, the four-point function reduces
to:
\begin{equation}
\begin{split}
\bigl<T_{1} T_{2} T_{3} T_{4}\bigr> =&\:\:\:\:\: \bigl<T_{1}
 T_{2}\bigr>\bigl<T_{3} T_{4}\bigr> \\& +
 \bigl<T_{1}T_{3}\bigr>\bigl<T_{2} T_{4}\bigr> \\&+ \bigl<T_{1} T_{4}\bigr>\bigl<T_{2} T_{3}\bigr> 
\label{eq:fourpt}
\end{split}
\end{equation}
For the four functions we defined in Sect.\
\ref{sec:definitions}, we find the following expressions:
\begin{align}
C^{1+3}(\theta) & = 3 \: C_{2}(0) \: C_{2}(\theta) \label{eq:13}\\
C^{2+2}(\theta) & = C_{2}(0)^2 + 2 \: C_{2}(\theta)^2 \\
C^{\mathrm{equi}}(\theta) &= C_{2}(0) \:C_{2}(\theta) + 2 \:C_{2}(\theta)^2\\
C^{\mathrm{rhomb}}(\theta) &= C_{2}(\theta) \:C_{2}(\theta') + 2
\:C_{2}(\theta)^2
\label{eq:rhombic}
\end{align}
In equation (\ref{eq:rhombic}) $\theta'$ denotes the length of the
longest axis of the rhombus, which is easily computed by spherical
trigonometry:
\begin{equation}
\cos \frac{\theta'}{2} = \frac{\cos \theta}{\cos \frac{\theta}{2}}
\end{equation} 

Note that the functions given by Eqs.\ (\ref{eq:13})-(\ref{eq:rhombic})
should be interpreted as expectation values. That is, the observed
four-point function should equal that given by Eq.\ (\ref{eq:fourpt})
to the same extent that the three-point function equals zero. Thus, to
establish an acceptable distribution for these relations we again utilize our
Monte-Carlo simulation set.
We compare the simulated ensemble of functions to those predicted by
Eqs.\ (\ref{eq:13})--(\ref{eq:rhombic}): for each realization in the ensemble, we
compute the four-point function predicted by the two-point function relation
above, and then evaluate the difference between this predicted
and the observed four-point function. From these differences we 
generate 68\% and 95\% confidence intervals. Finally, the procedure is
repeated for the DMR maps, and the results are compared to the
derived confidence intervals.

This procedure has one major advantage compared to the
one described in Sect.\ \ref{sec:measurements}: the power spectrum only
mildly affects the result. That is, the two most important contributions to the
analysis come from the map itself, in the form of a
two-point and a four-point function. The assumed power spectrum is
only used for estimating the acceptable \emph{deviations}, not the
overall shape. Therefore, this procedure provides a more direct
test for Gaussianity than the previous one.

\begin{table*}[!t]
\centering
\begin{tabular}{|c|l|cccc|cccc|c|}\hhline{*{11}{-}}  
\multicolumn{2}{|c|}{}&\multicolumn{4}{c|}{53 GHz maps}& \multicolumn{4}{c|}{90 GHz maps} &
\\ \hhline{~~*{9}{-}}
\multicolumn{2}{|c|}{}&$A$& $B$ & $\frac{\mathrm{A+B}}{2}$ & $\frac{\mathrm{A-B}}{2}$&$A$& $B$ &
$\frac{\mathrm{A+B}}{2}$ & $\frac{\mathrm{A-B}}{2}$ & Co-added \\ \hline
&Pseudo-collapsed 1+3 point &0.73 & 0.69 & 0.32 & 0.17 & 0.81 & 0.51 & 0.87 & 0.52 & 0.60\\
&Pseudo-collapsed 2+2 point&0.46 & 0.62 & 0.58 & 0.95 & 0.27 & 0.32 & 0.90 & 0.14 & 0.75\\
&Pseudo-collapsed equilateral four-point & 0.40 & 0.62 & 0.54 & 0.79 & 0.92 & 0.61 & 0.57 & 0.31 & 0.79\\
\hspace*{1mm}
\begin{rotate}{90}
\tiny
Uncorrected
\normalsize
\end{rotate}
&Rhombic four-point&0.49 & 0.98 & 0.60 & 0.17 & 0.50 & 0.94 & 0.75 & 0.09 & 0.76\\\hline
&Pseudo-collapsed 1+3 point &0.87 & 0.76 & 0.35 & 0.17 & 0.86 & 0.61 & 0.81 & 0.52 & 0.71\\
&Pseudo-collapsed 2+2 point &0.52 & 0.52 & 0.58 & 0.95 & 0.39 & 0.38 & 0.92 & 0.14 & 0.73\\
&Pseudo-collapsed equilateral four-point&0.54 & 0.71 & 0.69 & 0.79 & 0.94 & 0.79 & 0.74 & 0.31 & 0.81\\
\hspace*{1mm}
\begin{rotate}{90}
\tiny
\hspace*{0.3mm}
Corrected
\normalsize
\end{rotate}
&Rhombic four-point&0.50 & 0.98 & 0.67 & 0.17 & 0.58 & 0.94 & 0.80 & 0.09 & 0.87\\\hline
\end{tabular}
\caption{Results for the modified $\chi^2$ statistic. The numbers
refer to the fraction of simulations with $\chi^2$ values higher than
the corresponding \emph{COBE}-DMR maps. The upper half shows the results for
the uncorrected maps, while Galactic emission has been corrected for in
the lower half.}
\label{tab:red_chisq}
\end{table*}

The results are shown for the co-added map in
Fig.\ \ref{fig:reduced}. The observed function lies well 
within the confidence regions about the predicted function 
for all four cases. 
For a more quantitative measure of the perceived agreement,
we define a $\chi^2$ statistic, incorporating the new degree of freedom provided
by the predicted four-point function by simply replacing the average
correlation function with that new function:
\begin{equation}
\chi^2 = \sum_{\alpha \beta} (D_{\alpha} - D_{\alpha}^{\mathrm{pred}})
(\mathbf{M}^{-1})_{\alpha \beta} (D_{\beta} - D_{\beta}^{\mathrm{pred}})
\end{equation}
where
\begin{equation}
\mathbf{M}_{\alpha \beta} = \frac{1}{N} \sum_{i} (S^{i}_{\alpha}
- S_{\alpha}^{i,\mathrm{pred}}) (S^{i}_{\beta} - S_{\beta}^{i,\mathrm{pred}})
\end{equation}
The meaning of each symbol is the same as in Eqs.\ (\ref{eq:chisq}) and
(\ref{eq:m_mat}). 

Table \ref{tab:red_chisq} summarizes the results, which
again support the hypothesis that the DMR sky maps are consistent
with a scale-invariant cosmological model with Gaussian initial
fluctuations.

\section{Conclusions}

By performing Monte Carlo-simulations we have studied the statistical
properties of the \emph{COBE}-DMR 53 GHz and 90 GHz channels. The
basic ingredients for this analysis were various $N$-point correlation functions,
and, in particular, four different four-point functions which have been
presented for the first time.
We have additionally taken advantage of a result from statistical theory, 
relating all even $N$-point functions to reductions in terms
of the two-point function. 
This allowed us to define a test for Gaussianity in which the assumed power spectrum
only plays a secondary role. This test could therefore prove better
suited for situations in which we do not have access to the optimal
power spectrum.

Comparison of the DMR $N$-point correlation functions
with the Monte-Carlo ensemble indicates 
no evidence for possible non-Gaussian behavior, in agreement
with the earlier analysis of Kogut et al. (\cite{kogut}).
Furthermore, the agreement between the observed DMR functions and the simulated
ensembles also supports the validity of our model assumptions, namely
that of a scale-invariant power law model for the anisotropies, and
uncorrelated noise. 

On the other hand, the excellent agreement between the simulated and
the observed correlation functions poses an intriguing problem: tests of Gaussianity  
based on a harmonic analysis of the DMR data -- the bispectrum work
of Ferreira et al. (\cite{ferreira}) and trispectrum results of
Kunz et al. (\cite{kunz}) -- show compelling evidence for
non-Gaussian features (although these have subsequently been
associated with systematic artifacts in the DMR data by Banday et al.
\cite{bandaya}), while tests based on real-space 
high-order statistics such as those presented here do not.
The resolution of such apparently contradictory results is most likely
rather mundane: the source of the non-Gaussian signal was found to be 
strongly located at the multipole order $l=16$. 
Since the correlation functions are by definition (weighted) 
averages over the full multipole range, the reduced sensitivity to
this type of non-Gaussian structure is certainly not unexpected.

\begin{acknowledgements}
We acknowledge use of the HEALPix software and analysis package
for deriving the results in this paper. H.K.E. acknowledges
useful discussions with Per B. Lilje.
\end{acknowledgements}

\end{document}